\documentstyle[12pt,html,epsf,psfig]{article}

\begin{document}
\title{MATTERS OF GRAVITY, The newsletter of the APS Topical Group on 
Gravitation}
\begin{center}
{ \Large {\bf MATTERS OF GRAVITY}}\\ 
\bigskip
\hrule
\medskip
{The newsletter of the Topical Group on Gravitation of the American Physical 
Society}\\
\medskip
{\bf Number 24 \hfill Fall 2004}
\end{center}
\begin{flushleft}

\tableofcontents
\vfill
\section*{\noindent  Editor\hfill}

Jorge Pullin\\
\smallskip
Department of Physics and Astronomy\\
Louisiana State University\\
Baton Rouge, LA 70803-4001\\
Phone/Fax: (225)578-0464\\
Internet: 
\htmladdnormallink{\protect {\tt{pullin@phys.lsu.edu}}}
{mailto:pullin@phys.lsu.edu}\\
WWW: \htmladdnormallink{\protect {\tt{http://www.phys.lsu.edu/faculty/pullin}}}
{http://www.phys.lsu.edu/faculty/pullin}\\
\hfill ISSN: 1527-3431
\begin{rawhtml}
<P>
<BR><HR><P>
\end{rawhtml}
\end{flushleft}
\pagebreak
\section*{Editorial}

Given that relativity is about to be 100 years old, we 
decided to add a section ``100 years ago'' where we will
reprint (if copyright allows) famous papers from 100 years
ago. We start with a paper by Lorentz from 1904.

I want to encourage the readership to suggest
topics for articles in MOG. In the last few issues articles were
solicited by myself. This is not good for keeping the newsletter
balanced. Either contact the relevant correspondent or me directly.

The next newsletter is due February 1st. All issues are available in
the WWW:\\\htmladdnormallink{\protect
{\tt{http://www.phys.lsu.edu/mog}}} {http://www.phys.lsu.edu/mog}\\

The newsletter is  available for
Palm Pilots, Palm PC's and web-enabled cell phones as an
Avantgo channel. Check out 
\htmladdnormallink{\protect {\tt{http://www.avantgo.com}}}
{http://www.avantgo.com} under technology$\rightarrow$science.

A hardcopy of the newsletter is distributed free of charge to the
members of the APS Topical Group on Gravitation upon request (the
default distribution form is via the web) to the secretary of the
Topical Group.  It is considered a lack of etiquette to ask me to mail
you hard copies of the newsletter unless you have exhausted all your
resources to get your copy otherwise.  

If you have comments/questions/complaints about the newsletter email
me. Have fun.

\bigbreak
   
\hfill Jorge Pullin\vspace{-0.8cm}
\parskip=0pt
\section*{Correspondents of Matters of Gravity}
\begin{itemize}
\setlength{\itemsep}{-5pt}
\setlength{\parsep}{0pt}
\item John Friedman and Kip Thorne: Relativistic Astrophysics,
\item Bei-Lok Hu: Quantum Cosmology and Related Topics
\item Gary Horowitz: Interface with Mathematical High Energy Physics and
String Theory
\item Beverly Berger: News from NSF
\item Richard Matzner: Numerical Relativity
\item Abhay Ashtekar and Ted Newman: Mathematical Relativity
\item Bernie Schutz: News From Europe
\item Lee Smolin: Quantum Gravity
\item Cliff Will: Confrontation of Theory with Experiment
\item Peter Bender: Space Experiments
\item Riley Newman: Laboratory Experiments
\item Warren Johnson: Resonant Mass Gravitational Wave Detectors
\item Stan Whitcomb: LIGO Project
\item Peter Saulson: former editor, correspondent at large.
\end{itemize}
\section*{Topical Group in Gravitation (GGR) Authorities}
Chair: Jim Isenberg; Chair-Elect: Jorge Pullin; Vice-Chair: 
\'{E}anna Flanagan;
Secretary-Treasurer: Patrick Brady; Past Chair: John Friedman;
Delegates:
Bei-Lok Hu, Sean Carroll,
Bernd Bruegmann, Don Marolf, 
Gary Horowitz, Eric Adelberger.
\parskip=10pt
\vfill
\pagebreak

\section*{\centerline {
Message from the Chair}}
\addtocontents{toc}{\protect\medskip}
\addtocontents{toc}{\bf GGR News:}
\addcontentsline{toc}{subsubsection}{\it  
Message from the Chair, by Jim Isenberg}
\begin{center}
Jim Isenberg, University of Oregon
\htmladdnormallink{jim-at-newton.uoregon.edu}
{mailto:jim@newton.uoregon.edu}
\end{center}
Albert Einstein was born $125 \frac{1}{2}$ years ago in Ulm, Germany. A
little over 25 years later, Einstein published three epic papers which set
the course for much of theoretical physics for the twentieth century (and
beyond). One of them introduces special relativity, one of them proposes a
new ``quantum" model for understanding the photoelectric effect, and the
last describes a statistical mathematical model for understanding Brownian
motion.

Einstein's may be the most familiar face on the planet, and unlike his
main competitors (Osama bin Laden, Saddam Hussein, Mao, Castro, and
Michael Jackson), just about everyone likes him. If you tell someone that
you work on Einstein stuff, you get a much friendlier reaction than if you
say that you study differential equations or do LIGO data analysis.
Einstein has been used successfully to sell computer supplies, bagels,
nuclear weapons, national aspirations, IQ tests, healthcare, and board
games. 

This coming year, we should capitalize on our Einstein connection and use
him to sell gravitational physics, both within the scientific community
and out in public. The UN is helping: With Einstein in mind, it has
officially declared 2005 to be the ``International Year of Physics" (2004
is the ``International Year of Rice"). The APS is responding: It has a
number of programs planned, some kicking off this October, and it has
encouraged us to organize a special evening of talks on gravitational
physics at next April's meeting.
The public news media will likely be interested in Einstein and the
centennial of the 1905 papers as well; and if our local  media isn't, we
should tell them about it. In doing so, we can convey to them and to the
public some of the joy and satisfaction we get in studying general
relativity and gravity. And we can tell people about how dramatically this
area of scientific research has developed in recent years. The Speaker's
Bureau, which has been organized through the GGR in cooperation with other
groups, should help with this public mission. Its goals and operation are
described by Richard Price elsewhere in this issue of MOG. 

A key purpose of the topical group is to tell people about gravitational
physics, get them interested in it, and encourage fresh talent to enter
the field. Einstein, though over 125 years old, should be a big help this
year in achieving this goal.

\section*{\centerline {
We hear that...}}
\addcontentsline{toc}{subsubsection}{\it  
We hear that..., by Jorge Pullin}
\begin{center}
Jorge Pullin, LSU
\htmladdnormallink{pullin-at-lsu.edu}
{mailto:pullin@lsu.edu}
\end{center}

$\bullet$ Bryce DeWitt is this year's winner of the {\it Einstein Prize}
of APS.

Hearty congratulations!

\vfill
\eject

\section*{\centerline {
THE TGG WYP Speakers Program}}
\addcontentsline{toc}{subsubsection}{\it  
THE TGG WYP Speakers Program, by Richard Price}
\begin{center}
Richard Price, University of Texas at Brownsville
\htmladdnormallink{rprice-at-physics.utah.edu}
{mailto:rprice@physics.utah.edu}
\end{center}
Your Executive Committee felt that the topical group has a special
relationship to the 2005 International Year of Physics.  Einstein is the
spur for WYP, and Einstein is ours. We sponsor the Einstein prize, don't
we? And Einstein, after all, would have been a TGG member. As Chairman Jim
points out in his message, there will be enhanced interest in Einstein,
and in Einstein physics in the WYP. Our topical group has the experts in
the physics, so the decision was made, almost two and a half years ago, to
share our expertise with the world; the TGG WYP Speakers program was
started.

The TGG WYP task force, Chairman Jim, Jennie Traschen and myself, set
up a program to connect speaker requests with appropriate
speakers. (Speakers on the right subject who can speak at the right
technical --- or nontechnical -- level.)  The program, costing
nothing, is well within the TGG budget. The administration is carried
out by a staff member, Danka Mogilska, of the University of Texas at
Brownsville (my new home). The requesting group sends in, by mail,
email, or webform, a request with all the information, and Danka looks
for a suitable speaker close enough so that negligible travel is
involved. If a candidate speaker is found, then the speaker is invited
to contact the requesting group. In this way, the inconvenience to the
speakers is minimized.

It appears to others also now, that such a speakers program is a good
idea. The WYP effort of the Forum on the History of Physics, under the
leadership of Virginia Trimble, is quite naturally very interested in
the Einstein connection in the WYP. A month ago, they joined with us
in the Speakers program. We can now offer experts on Einstein as well
as on his physics. Very recently we also established a tie to the
Division of Astrophysics. This is so recent, in fact, that details
have not been worked out. It is likely that they will provide a
different kind of service (in which they pay travel funds!), but it
will be useful to have a central administration for all these
speakers.

Early on, the Executive Committee selfishly decided that our priority
in the Speakers program would be undergraduate institutions, colleges
that have potential gravity physics grad students. It appears at
present that prioritization is necessary. Requests began rolling in at
the end of July, and there are many we will not be able to fill.

You can see our website by going to the APS website (www.aps.org)
and following the WYP link to the Speakers Program, or you can
go directly to

\htmladdnormallink{        
        http://www.phys.utb.edu/WYPspeakers/REQUESTS/howto.html}
{http://www.phys.utb.edu/WYPspeakers/REQUESTS/howto.html}

It should be a long, but interesting year.
\vfill
\eject

\section*{\centerline {
Gravity Probe B is launched from Vandenberg AFB}}
\addtocontents{toc}{\protect\medskip}
\addtocontents{toc}{\bf Research Briefs:}
\addcontentsline{toc}{subsubsection}{\it  
Gravity Probe B is launched, by Bill Hamilton}
\begin{center}
Bill Hamilton, Louisiana State University
\htmladdnormallink{hamilton-at-phys.lsu.edu}
{mailto:hamilton@phys.lsu.edu}
\end{center}

April 20, 2004 marked the end of the beginning phases for the Gravity
Probe B experiment with the launch of the experiment aboard a Delta II
rocket from Vandenberg AFB.  This experiment has had an
extraordinarily long gestation period as initial discussions between
Bill Fairbank, Leonard Schiff and Bob Cannon at the Stanford men's
swimming pool in 1958 combined with Francis Everitt joining the effort
in 1962 culminated in the launch in 2004.  Now, as Francis said after
the launch, the real work begins.

Gravity Probe B is an elaborate package that is designed to directly
demonstrate two effects.  The first, the geodetic effect, measures the
precession of a gyroscope orbiting the earth.  If a perfect gyroscope
is placed in a free falling orbit around a massive object the angular
momentum of the gyroscope will be connected with the orbit through
parallel transport.  Since the space-time is curved the geometry of
the orbit is not Euclidean.  As Everitt put it at the press conference
before the launch: ``If the orbit were perfectly circular its
circumference would be just a couple of centimeters less than $\pi$
times the orbital diameter."  Thus the gyro angular momentum exhibits
a precession in the plane of the orbit.  The second effect, the
Lense-Thirring precession, results from the dragging of the orbital
inertial frame by the rotating earth.  This precession can be made to
be perpendicular to the geodetic precession if the orbit is polar.

The predicted precessions are very small.  The geodetic precession is
calculated to be 6614.4 milliarcseconds per year; the frame dragging
precession is supposed to be 40.9 milliarcseconds per year.  These
incredibly small precession angles required the development of many
new technologies.  The experiment's orbit has to be continuously
corrected for drag, i.e. it must be a zero g orbit.  The gyroscopes
need to be perfect spheres to eliminate the torques due to
gravitational gradients and techniques had to be developed to make
them and to measure their orientation without inducing torques.  A
technique needed to be developed to maintain a fixed direction in
inertial space and a whole new type of telescope was developed to
sense this direction and couple it to the gyroscopes.  All of this
means that literally hundreds of people have been involved in, or have
contributed to, various aspects of the GPB spacecraft.

Not everyone who came or who tried to come was able to see the launch
because, in the best tradition of Gravity Probe B, there were delays.
The original launch of 6 December 2003 had to be postponed to rework
some grounding problems in the experimental control unit.  It was then
scheduled for Saturday, 17 April 2004.  Most of us who had tickets
already were able to change them to the later date.  NASA had been
very explicit that we had to have official letters of invitation to
attend and they were reissued for the later date.  On 7 April we were
informed that the launch was postponed for two days, until Monday, 19
April because of a short circuit in the launch tower.  Everyone I
talked to said that they just decided that it was ``this time for sure"
and that they weren't going to change their tickets.

The launch activities were a virtual Woodstock of old gravitational
physicists.  NASA's official visitor's headquarters was in the small
town of Buellton.  It was there that the press conferences were staged
and it was from there that buses were scheduled to take people out to
the launch site.  A number of people, many of those from Stanford,
stayed nearer to the Vandenberg gates at Lompoc.  I saw people that I
hadn't seen in 15 or 20 years.  There were several NASA briefings at
the Marriott in Buellton.  The powers-that-be had been thoughtful
enough to invite the children of Leonard Schiff and Bill Fairbank to
see the launch and to participate in the briefings.

On Monday morning, 19 April, a vast fleet of buses appeared in front
of the Marriott and hauled us all out to the official viewing site.
We discovered that we couldn't see the rocket on the launch site at
all: it was hidden behind a small hill.  We did have television
monitors underneath a big tent and loud speakers connected to launch
control but there was no view of the launch tower.  A number of us
with binoculars and absolutely no knowledge speculated about just what
things visible on the horizon might be the launch tower.  We all took
pictures of the horizon and each other and eagerly listened to the
countdown until t= -3 when it was announced that the winds aloft were
too high and the launch would be postponed until tomorrow: Tuesday 20
March.

The delay was unfortunate for a number of the prospective viewers
because of class commitments or plane reservations.  Several had to
leave.  Some of us with rental cars decided to scout the area to see
if there was a better viewing spot.  We found a place behind the
weather station where the launch tower was in full view at a distance
of about 4 km and we were told that the weather station personnel
would hook up a loud speaker to listen to mission control.  Talking to
old time Vandenberg personnel at a party for the invitees hosted by
Brad Parkinson reinforced the desirability of this spot.

The next morning we drove to our newfound viewing site and, by the
time of the launch, were sharing it with about 250 other people
including many past and present members of the Stanford team.  Rumors
were passed about the reason for the previous day's delay, the
prevalent one being that there had been a checksum error in loading
the wind profile into the guidance system.  Knowledgeable people spoke
about the reasons for various gas ventings from the rocket.  Finally
when the count got to t=0 there was a tremendous cloud of smoke from
the solid rocket boosters and the vehicle began to rise above the
tower.  It took fewer than 3 minutes until the rocket was completely
out of sight, leaving only a trail of smoke in the sky.  My main
impression from the launch was that the brightness of the rocket's
flame was much greater than anything I could have imagined from what I
had seen of launches on television.  The other impression, of course,
was of how many people's work had gone into this project to get it
this far and how much more there still was to do.

As an experimentalist I am sometimes accused of being over-awed by
technology.  GPB is, to me, much more than a technological
demonstration.  It is an attempt to measure, directly, things that we
now believe we know.  We didn't know these things when the experiment
started.  We still don't have direct measurements of them.  This
experiment is a direct measurement in an environment we believe we
understand.  I can only be amazed at the persistence of the Stanford
team and Francis Everitt.  It took more than 40 years to develop the
technology and methodology to make these measurements.  In a year we
should have a firmer foundation for our theoretical speculations.

At this writing the experiment is working well.  To check on its
progress look at 
\htmladdnormallink{http://einstein.stanford.edu}
{http://einstein.stanford.edu}.

\vfill
\eject

\section*{\centerline {
Questions and progress in mathematical general relativity}}
\addcontentsline{toc}{subsubsection}{\it  
Questions and progress in mathematical general relativity, by Jim Isenberg}
\begin{center}
Jim Isenberg, University of Oregon
\htmladdnormallink{jim-at-newton.uoregon.edu}
{mailto:jim@newton.uoregon.edu}
\end{center}

Maxwell's equations are vitally important for describing the physical
universe but raise very few serious questions mathematically. There is
considerable mathematical interest in the global behavior of wave maps but
there is little evidence that wave maps play a direct role in modeling
important physical phenomena. One of the very few field equation systems
which is both a vital tool for modeling physics and  a very rich source of
serious research problems in math is that of Einstein. 

{}From a mathematical point of view, the study of Einstein's equations lies
in the realm of geometric analysis, which analyzes PDEs (partial
differential equations) that involve structures from differential
geometry. For some of the PDEs of geometric analysis, such as the minimal
surface equation and the harmonic map equation, the PDE system is elliptic
(potential-like); for others, like Ricci flow and mean curvature flow, the
PDEs are parabolic (heat-like) ; while for still others, such as wave maps
and Yang-Mills, the PDEs are hyperbolic (wave-like). Interestingly in the
case of Einstein's equations, there are aspects of all three types: the
Einstein constraints are essentially elliptic, the full system is
essentially hyperbolic, and when certain ansatze are applied  (such as
that of  Robinson-Trautman) parabolic analysis is called for. As a
consequence of this feature, the mathematical study of the Einstein
equation system intersects with a much wider variety of areas in geometric
analysis than does that of most other geometric analysis PDEs.

Rather than attempting any sort of an overview, I want to focus now on a
number of mathematically interesting questions which have arisen from the
study of Einstein's equations. In almost all of them there has been
substantial progress in recent years. 

\textbf{Positive Mass and the Penrose Inequality}: It is relatively
straightforward to define an integral quantity which measures the mass of
an isolated gravitational system. One of the most intensely pursued
questions of mathematical GR during the 1960's and 1970's was whether one
could prove that such a quantity is either positive or zero, and zero if
and only the spacetime is flat. A theorem to that effect was proven
finally in 1979 by Schoen-Yau [1] and in 1981 by Witten
[2] independently. Since then, interest has shifted to showing that
if a spacetime contains black holes, then the mass must be greater than or
equal to a quantity involving the square root of the areas of the horizons
(the ``Penrose Inequality"). Two very  innovative proofs of this
conjecture, \textit{for the time symmetric case} (initial data with
vanishing extrinsic curvature), have been produced in recent years by Bray
[3] and by Huisken-Ilmanen [4].  One would very much like
to extend this work to the non time symmetric case; this appears to be
very difficult. We note that in the time symmetric case, the question can
be analyzed purely in terms of Riemannian geometry. This is not true of
the more general case.

\textbf{Shielding Gravitational Effects}: In Maxwell's theory, a conductor
can be used to mask from the view of outside observers many of the details
of a charge configuration. Since there are no known negative masses in
gravitational  physics, it has long been thought that there are no such
conductor-like objects for shielding from external view the details of a
mass or gravitational field configuration. The elliptic character of the
Einstein constraints has reinforced this conjecture. Recent work on the
gluing of solutions of the Einstein constraints now belies this idea. In
particular the work of Corvino-Schoen [5,6] shows that
for essentially any asymptotically Euclidean initial data for Einstein's
equations, one can cut out the region outside of some ball and replace it
by a smooth extension which, some finite distance out, is exactly data for
the Schwarzschild or Kerr solution. The interior details are lost to
observers who are far enough away. The gluing results of
Chrusciel-Isenberg-Mazzeo-Pollack [7,8] allow
other quite surprising joins of disparate sets of initial data. One finds
as a consequence of their work that a given region may contain an
arbitrary number of wormholes without affecting at all the gravitational
fields some distance away from the region. All of these gluing results
exploit the underdetermined nature of the Einstein constraint equations
(more fields than equations). There is likely much more that can be done
to exploit this feature. In particular, one would like to know if it
allows one to always extend a given solution of the constraints on a ball
to an asymptotically Euclidean solution on $R^3$. This question plays a
role in the Bartnik approach to defining the ``quasilocal mass"of a given
region.

\textbf{Cosmic Censorship and the Nature of Singularities}: The
Hawking-Penrose singularity theorems indicate that singularities (in the
sense of geodesic incompleteness) generically occur in solutions of
Einstein's equations , but they say little about the nature of these
singularities. Roughly speaking, one expects the singularities to be
characterized either by curvature blowup or by the breakdown of causality
(marked by the formation of a Cauchy horizon). The Strong Cosmic
Censorship Conjecture (SCC) suggests that curvature blowup occurs
generically, and Cauchy horizons develop only in very special cases.
Proving SCC is a big challenge, requiring detailed knowledge of how
solutions evolve generically. There has, however, been progress in proving
the conjecture in limited families of solutions, defined by the presence
of symmetries. The most recent work in this direction, done by Ringstrom
[9] proves that SCC holds for the class of Gowdy solutions on the
torus. One of the key steps used by Ringstrom (as well as his
predecessors) is the verification that the Gowdy solutions are velocity
dominated near the singularity. This approach may be useful for certain
less specialized families of spacetimes which also appear to be velocity
dominated, like the polarized solutions with $U(1)$ symmetry,  While more
general families of solutions are likely to not have this property,
numerical studies indicate that they may be oscillatory in the BKL sense,
and the challenge now is to verify this, and use it to prove SCC. 

The very recent work of Dafermos [10] on the stability of the Cauchy
horizons found in the Reissner-Nordstrom solutions (spherically symmetric
charged black holes) is also relevant to the question of SCC.
Interestingly, he finds stability for certain differentiability classes of
perturbations, but not for others. Further development of this approach
should be very useful

The Weak Cosmic Censorship Conjecture (WCC), which is \textit{not} a
consequence of SCC, concerns a very different question: If a singularity
develops as a result of collapse in an asymptotically flat solution, does
an event horizon generally develop and cover it from view by observers at
infinity? As with SCC, WCC is a conjecture concerning the behavior of
\textit{generic} solutions, not every solution. While interest in this
question is strong, there have not been any important recent results
relevant to WCC.

\textbf{Long Time Behavior of Solutions and Stability}: For any nonlinear
hyperbolic PDE system with a well-posed Cauchy problem, one of the
questions of primary interest is whether one can characterize those
initial data sets for which solutions exist for all (proper) time. Since
singularities do generically develop, one cannot expect long time
existence in both time directions; but there is no reason to expect
singularities in both directions. For a system as nonlinear as Einstein's
equations, long time existence is  a very difficult problem. As a first
step towards its study, one approach is to investigate the stability of
long time existence about solutions like Minkowski space for which it is
known to hold. Some years ago, Christodoulou and Klainerman [11]
showed that indeed Minkowski space is stable in this sense. The recent
works of Klainerman-Niccolo [12] and of Lindblad-Rodnianski
[13] prove roughly the same result, but with stronger control of
the asymptotic properties of the spacetimes, with different choices of
gauge, and with the use of techniques which are much simpler. One would
like to extend these results to Schwarzschild and to Kerr, but there is to
date no progress in this direction.

Andersson and Moncrief  [13] have studied the stability of long
time existence for a different sort of spacetime: They have proven
stability for the expanding, spatially compact spacetimes one obtains by
compactifying the constant negative curvature hyperboloids in the future
light cone of the origin in Minkowski spacetime. It is not known what the
spacetime developments of the perturbed data do in the contracting
direction toward the singularity, but in the future direction, they find
that in fact the perturbed spacetimes asymptotically approach the flat
spacetimes from which they have been perturbed. Similar stability results
have been obtained by Choquet-Bruhat and Moncrief [14] for
expanding $U(1)$ symmetric spacetimes, and it is intriguing to speculate
that for rapid enough expansion, stability of long time existence is a
general feature. 

Another line of research related to the long time behavior of solutions
focuses on determining which sets of asymptotically Euclidean initial
data develop into spacetimes which can be conformally compactified so as
to include the familiar ``scri" structure at null infinity. Until
recently, the only spacetimes known to have this complete structure were
stationary. Combining the gluing work of Corvino-Schoen discussed above
with Friedrich's [15] work on the generation of scri from
sufficiently small initial data on an asymptotically hyperbolic spacelike
surface, one now knows that there are non stationary, radiating solutions
with a complete scri. The recent work of Kroon [16] shows that the
nature of scri can, for general spacetimes, become quite complicated
(including ``polyhomogeneous" behavior.)

Finally, we note that one approach towards obtaining long time existence
for a hyperbolic system is to try to reduce the level of  regularity
(number of derivatives) needed to prove well-posedness, and then find a
conserved norm compatible with that level of regularity.
Klainerman-Rodnianski [17] and Smith-Tataru [18] have
focused on this approach, and have managed to prove well-posedness for
data with the metric in the Sobolev space $H^{2+\epsilon}$, and the
extrinsic curvature in $H^{1+\epsilon}$. The first group is working very
hard  to remove the $\epsilon$.
\\
\\

There are a number of other very interesting areas of mathematical study
of the Einstein equations, including work to establish an initial
value-boundary value formulation of the system, efforts to understand and
parameterize non constant mean curvature solutions of the Einstein
constraint equations, attempts to obtain an analytical understanding of
the critical solutions found in Choptuik's numerical studies of
gravitational collapse, and continued studies of the ``static stars are
spherical" conjecture.  There is every expectation that the recent record
of progress in this area will continue for some time to come.
\\
\\

\parskip=5pt
{\bf References:}

[1]  R. Schoen and S.T. Yau, {\em On the proof of the positive
mass conjecture in general relativity}, Comm. Math. Phys. {\bf 65} (1979),
45-76.

[2]  E. Witten, {\em A simple proof of the positive energy
theorem}, Comm. Math. Phys. {\bf 80} (1981), 381. 

[3]  H. Bray, {\em Proof of the Riemannian Penrose inequality
using the positive mass theorem}, J. Diff. Geom. {\bf 59} (2001), 177-267.

[4]  G. Huisken and T. Ilmanen, {\em The Inverse mean curvature
flow and the Riemannian Penrose inequality}, J. Diff. Geom. {\bf 59}
(2001), 353-437.

[5] J. Corvino,{\em Scalar curvature deformation and a gluing
construction for the Einstein constraint equations}, Comm. Math. Phys.
{\bf 214} (2000), 137-189.

[6] J. Corvino and R. Schoen, {\em On the asymptotics for the
vacuum Einstein constraint equations}, 
\htmladdnormallink{gr-qc/0301071}
{http://arxiv.org/abs/gr-qc/0301071}.

[7] J. Isenberg, R. Mazzeo, and D. Pollack, {\em Gluing and
wormholes for the Einstein constraint equations}, Comm. Math. Phys. {\bf
231} (2002), 529-568.

[8] P. Chrusciel, J. Isenberg, and D. Pollack, {\em Initial
data engineering}, \htmladdnormallink{gr-qc/0403066}
{http://arxiv.org/abs/gr-qc/0403066}.

[9] H. Ringstrom, {\em Asymptotic expansions close to the
singularity in Gowdy spacetimes}, Class. Quan.  Grav. {\bf 21} (2004),
S305-S322.

[10] M. Dafermos, {\em The interior of charged black holes and
the problem of uniqueness in general relativity}, 
\htmladdnormallink{gr-qc/0307013}
{http://arxiv.org/abs/gr-qc/0307013}.

[11] D. Christodoulou and S. Klainerman, {\em The Global
nonlinear stability of the Minkowski space}, Princeton Math Series
{\bf41}, Princeton U. Press, Princeton, N.J., (1993). 

[12] S. Klainerman  and F. Nicolo, {\em Peeling properties of
asymptotically flat solutions to the Einstein vacuum equations}, Class.
Quan. Grav. {\bf 20} (2003), 3215-3257.

[13] H. Lindblad and I. Rodnianski, {\em Global existence for
the Einstein vacuum equations in wave coordinates}, 
\htmladdnormallink{math.AP/0312479}
{http://arxiv.org/abs/math.AP/0312479}.

[14] L. Andersson and V. Moncrief, {\em Future complete vacuum
spacetimes}, \htmladdnormallink{gr-qc/0303045}
{http://arxiv.org/abs/gr-qc/0303045}.

[15] Y. Choquet-Bruhat and V. Moncrief, {\em Future complete
Einsteinian spacetimes with $U(1)$ isometry group}, Ann. H. Poincare {\bf
2} (2001), 1007-1064. 

[16] H. Friedrich {\em On the existence of n-geodesically
complete or future complete solutions of the Einstein field equations with
smooth asymptotic structure}, Comm. Math. Phys. {\bf107} (1986), 587-609.

[17] J. Kroon, {\em Polyhomogeneous expansions close to null and
spatial infinity}, \htmladdnormallink{gr-qc/0202001}
{http://arxiv.org/abs/gr-qc/0202001}.

[18] S. Klainerman and I. Rodnianski, {\em Rough solutions of
the Einstein vacuum equations},  C. R. Math. Acad. Sci. Paris {\bf 334}
(2002), 125-130.

[19] H. Smith and D. Tataru, {\em Sharp local well-posedness
results for the nonlinear wave equation}, preprint (2001).

\parskip=10pt

\vfill
\eject

\section*{\centerline {
Summary of recent preliminary LIGO results}}
\addcontentsline{toc}{subsubsection}{\it  
Summary of recent preliminary LIGO results, by Alan Wiseman for the LSC}
\begin{center}
Alan Wiseman, University of Wisconsin-Milwaukee for the LIGO Science Collaboration
\htmladdnormallink{agw-at-uwm.edu}
{mailto:agw@uwm.edu}
\end{center}

\bigskip

The LIGO Lab and the LIGO Scientific Collaboration (LSC) continue
to interweave detector commissioning and data taking with data
analysis and the presentation and publication of scientific
results.  In the Spring 2003 issue of {\it Matters of Gravity},
Gary Sanders summarized the status of the LIGO detector
commissioning effort leading up to the first Science Run (S1)
and recapped the preliminary ``upper-limits" results that Albert
Lazzarini had reported at the AAAS meeting in Denver.  In last
Fall's issue, Stan Whitcomb reported on the completion of a
second LIGO data taking run, S2.  In the intervening year, the
LIGO Scientific Collaboration has published a sequence of five
major articles that culminate the work on the S1 data set [1-5],
as well as numerous conference proceedings and technical reports
[6].  We also completed another two-month science run (S3) in
early January of 2004. In this note,  I would like to briefly
recap the published S1 results, as well as summarize the
preliminary S2 and S3 results that were presented at the Denver
APS meeting in May and the Dublin GR17 meeting in July.

The data analysis effort within the LSC is currently divided
among four groups reflecting four distinct source types: the
Inspiral Upper Limits Group, the Stochastic Background Upper
Limits Group, the Pulsar (continuous waves) Upper Limits Group
and the ``Burst" Upper Limits Group.  When the groups were formed
some years ago, the qualifier ``upper-limits" was included in the
group name to reflect the fact that the sensitivity of the
instrument during the early running would likely lend itself to
only setting upper limits on flux strength and population models.
However, as the sensitivity of the detectors has improved, each
group has begun to set their sites on true detections, and thus the
use of the qualifier is falling by the way-side.  

However, the first article [1] to wind its way through both the internal
LSC review\footnote{Most of us in gravitational physics have
written papers in small collaborations and know the difficulties
of negotiating with our colleagues about the precise content of
the paper: imagine scaling this process to an author list with
over 300 people!} and the external peer-review process was not
an astrophysical paper originating from within  the
analysis groups. Rather, the first paper gave a detailed
description of the configuration and performance of the
LIGO detectors and the British-German GEO detector during   the
17 day S1 data run in August and September of 2002.  This
``detector" paper was then followed by ``upper-limits" papers from
each of the four search groups.  Although the final astrophysical
results in these four papers do not challenge any existing
theories, they do present complete analyses which show how to
search {\it real} data for small gravitational wave signals and
how to translate those searches into astrophysical limits.

\begin{figure}[h]
\psfig{file=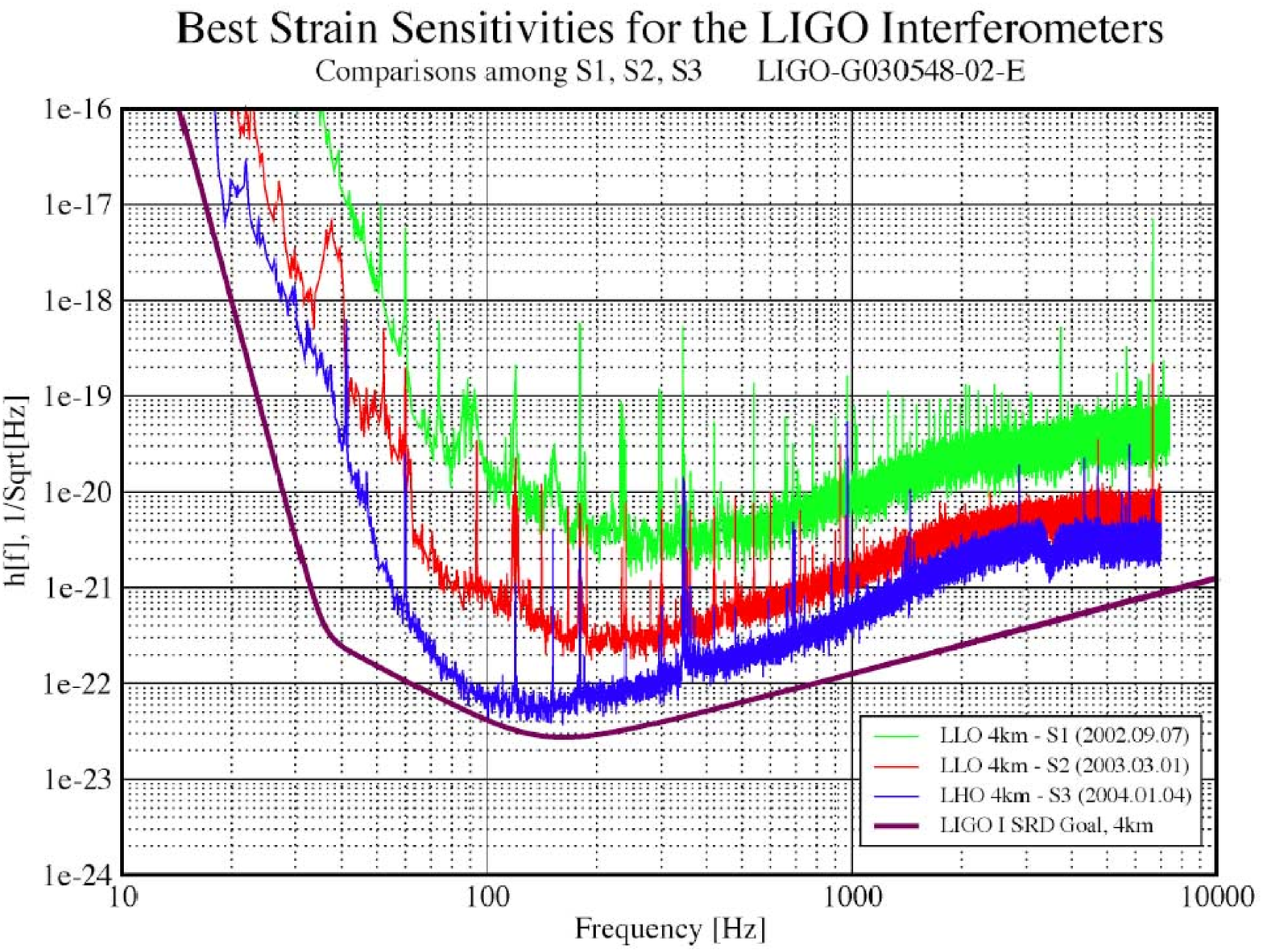,width=5.5in,height=4.75in} 
\caption{Best Strain Sensitivities for the LIGO Interferometers.
LHO refers to the Hanford Observatory. LLO refers to the
Livingston Observatory.
\label{f:strain} 
}
\end{figure}

The presentation of the preliminary S2 and S3 results at the APS and
GR meetings followed a pattern similar to the S1 publications: a
summary presentation describing the status of the detectors
was followed by a talk from each of the four analysis groups.
Although the results are summarized below, we invite everyone to
take a look at the vu-graphs that were presented [7].
The central feature of the summary talk was to
show the dramatic improvement in detector sensitivity over the
last few years.  Figure \ref{f:strain} shows that the commissioning efforts
are paying off, and the detector is nearing the design
sensitivity.

In the published S1 analysis [2], the Burst Group took on the
daunting task of limiting the rate and strength of poorly modeled
burst sources of gravitational waves. [It is hard to look for
something when you don't know what it looks like.]  The primary
result of this analysis was to quantify an excluded (or low probability) 
region in the rate versus signal strength plane. Although a
similar analysis is being repeated with the more sensitive S2 and
S3 data, the Burst Group has also added a new type of search to the
mix: a ``triggered" burst search. Some preliminary results from
this search were presented at the APS and the GR17 meetings.
During S2 an especially strong $\gamma$-ray burst (GRB 030329)
popped off nearby ($z=0.1685$).  The data from the two Hanford
detectors were cross correlated in a 180 second interval
surrounding the arrival time of the burst.  [Unfortunately the
Livingston and GEO detectors were off-line during the burst.]
This was compared to a threshold set by a similar analysis
conducted on data taken well away from the burst arrival time.
Although no gravitational wave burst was detected, a strain upper limit
of $h_{rss} \approx 6\times10^{-21} {\rm Hz}^{-1/2}$ was set.

The Inspiral Group searched for non-spinning binary neutron star
inspirals in 236 hours of S1 data [3]. Unfortunately, the
sensitivity of the instruments only allowed them to see inspirals
within a portion of the Milky Way Galaxy. The group used a
``loudest event" statistic to determine the upper limit on the
event rate of neutron star coalescences.  The published S1 paper gives
a  ninety-percent confidence limit on the event rate of binary
neutron star coalescences as
\begin{eqnarray}
R_{S1} < 1.7 \times 10^2 \; {\rm per \;year \; per \; MWEG}
\; . \nonumber
\end{eqnarray}
Here, MWEG means Milky Way Equivalent Galaxy.  During the S2 run
the instrument was considerably more sensitive and could see
beyond the Galaxy. For example, the Livingston Detector,
could reach M31. However, because the group is beginning to look
beyond simply setting upper limits and toward a possible detection,
they have modified their analysis pipeline accordingly.  Only
coincident data (when two or more interferometers were operating)
was included in their upper limit. This reduced the ``live time"
to about 355 hours and therefore reduced the upper limit that
might have been attained had they analyzed the ``singles" S2 data in
the same way they analyzed the S1 data.  Nevertheless, an
improved preliminary upper limit of 
\begin{eqnarray}
R_{S2} < 50 \; {\rm per \;year \; per \; MWEG} \nonumber
\end{eqnarray}
was obtained.

As the Inspiral Analysis Group moves forward, they will be
casting a broader net and include searches for binary black holes
with masses greater than $3 M_{\rm Sun}$.  Searching for these
signals presents a special challenge as neither the waveforms or
the population models are well known. They will also be looking
for inspiraling massive halo objects (MACHOs) with masses in the
range of $0.2 - 1.0M_{\rm Sun}$. The MACHO search presents a different
challenge as the template waveforms for these low mass systems
spend much longer in the LIGO sensitivity band than do the
neutron star binaries.

In analyzing the S1 data [4], the Pulsar Analysis Group used two very
different techniques to  perform a search aimed at setting an
upper limit on the ellipticity of PSR 1939+2134: a
Bayesian time-domain search and a frequency-domain search.
Several features of this search are worth noting.  First, a
directed search at a pulsar with known frequency and sky location
is less computationally intensive than a search for unknown
pulsars in a broad region of the sky or in a broad frequency
band.  Second,
during S1, both GEO and LIGO were operating, and therefore the
results combine information from both detectors.  Third, the fact
that two different analysis methods arrived at essentially the
same results instills confidence in the implementation of both
methods.  The result is a bound on the ellipticity of this
pulsar of $\epsilon_{S1} < 2.9\times 10^{-4}$.

Using the S2 data, the Pulsar Analysis  Group has now  applied
the ``time domain" search method to 28 known isolated pulsars with
good timing data.  In the case of PSR 1939+2134, this
analysis has improved the upper limit on the
ellipticity by about an order of magnitude. A full list of the
ellipticity bounds should be published soon. The Pulsar Analysis
Group also has ambitious plans for doing wide parameter space
searches to look for unknown pulsars.

A search through the S1 data for a stochastic background of
gravitational waves has also been completed and published
[5]. The basic idea is to cross-correlate data from
two detectors with the appropriate range of time lags. The result
for the Hanford-Livingston cross correlation is a bound on
$\Omega_{gw}h_{100}^2 < 23 \pm 4.6$ in the frequency band
$64-265$Hz, where $\Omega_{gw}$ is the energy density per
logarithmic frequency interval divided by the energy
density required to close the universe, and $h_{100}$ is the Hubble constant
in units of $100{\rm km/sec \; Mpc}$.  At the GR17 meeting, a preliminary
S2 result of $\Omega_{gw}h_{100}^2 < 0.018 (+0.007,-0.003)$ in the
frequency band $50-300$Hz was presented.  Based on sensitivity
estimates from the S3 noise curves, we expect to be able to place
a bound of roughly $5\times 10^{-4}$ in the same frequency band.

{\bf References:}
\begin{itemize}
\item 
{\it Detector Description and 
Performance for the First Coincidence Observations
between LIGO and GEO.} Nucl.Instrum.Meth. A517 (2004) 154-179.

\item {\it First upper limits from LIGO on
gravitational wave bursts}. Phys. Rev. D 69 (2004) 102001.

\item {\it Analysis of LIGO data for gravitational
waves from binary neutron stars}. Phys.Rev. D69 (2004) 122001.

\item {\it Setting upper limits on the strength of 
periodic gravitational waves using the first
science data from the GEO600 and LIGO detectors}.  Phys.Rev. D69 (2004) 082004.

\item {\it Analysis of First LIGO Science Data for
Stochastic Gravitational Waves}. Phys.Rev. D69 (2004) 122004.

\item In addition to [1-5], there have
also been numerous conference proceedings and technical
articles that have appeared.  Most can be found by searching for
the LIGO Collaboration at
\htmladdnormallink{
http://www.slac.stanford.edu/spires/hep/search/index.shtml}
{http://www.slac.stanford.edu/spires/hep/search/index.shtml}.

\item
\htmladdnormallink{http://www.ligo.org/}{http://www.ligo.org/}
 $\rightarrow$ Observational Results.
\end{itemize}

\section*{\centerline {
100 Years ago}}
\addcontentsline{toc}{subsubsection}{\it  
100 Years ago, by Jorge Pullin}
\parskip=3pt
\begin{center}
Jorge Pullin
\htmladdnormallink{pullin-at-lsu.edu}
{mailto:pullin@lsu.edu}
\end{center}
In 1904 H. A. Lorentz published the paper ``Electromagnetic phenomena in
a system moving with any velocity less than that of light'' in 
Proceedings of the Academy of Sciences of Amsterdam. The paper is 
reprinted in its English version in the Dover book ``The principle
of relativity''. A scan of this version is available in 

\htmladdnormallink{http://www.phys.lsu.edu/mog/100}
{http://www.phys.lsu.edu/mog/100}
\vfill
\eject

\section*{\centerline {
Einstein 125}}
\addtocontents{toc}{\protect\medskip}
\addtocontents{toc}{\bf Conference reports:}
\addcontentsline{toc}{subsubsection}{\it  
Einstein 125, by Abhay Ashtekar}
\parskip=3pt
\begin{center}
Abhay Ashtekar, PennState
\htmladdnormallink{ashtekar-at-gravity.psu.edu}
{mailto:ashtekar@gravity.psu.edu}
\end{center}
The German Physical Society celebrated Einstein's 125th birthday
in Ulm, his birthplace, through a conference. Since the membership
of the Society is very large ---according to the organizers, it is
the largest Physical Society worldwide--- they have to divide
their annual meetings into sections. This meeting brought together
sections on General relativity Gravitation, History of Physics and
Mathematical Physics.

The conference was inaugurated by the German President, Johannes
Rau on March 14th, the day Einstein was born. There was a
delightful evening lecture entitled ``Einstein's Impact on
Theoretical Physics of the 21st Century'' by Professor C.N. Yang.
Although he is now 81 and had flown to Germany from Hong Kong just
the day before, there was not the slightest trace of jet-lag or
fatigue. He started out by saying that in his view, Einstein was
by far the greatest scientist of the 20th century and
characterized his work by {\it Depth, Scope, Imagination,
Persistence, and Absolute Independence}. Then, in his usual crisp
style, he summarized the continuing impact of Einstein's creative
contributions, particularly, his ``obsessions with unified field
theory''. Professor Yang sharply disagreed with the view that this
work was misguided or futile. Specifically, he explained that
three key themes in contemporary physics originate in that work:
Geometrization of physics; non-linearity of laws of Nature; and
the role of topology in physics.

This lecture was followed by a surprise: a musical event featuring
Paul Einstein on violin and Siegfried R\"abblen on Piano. Paul, a
great grandson of Einstein's is a musician living in the south of
France and played on Einstein's violin. The piece was a Mozart
Sonata, K304, written in 1778. It is the only instrumental work
Mozart wrote in E-minor and its poignancy reflects Mozart's
reaction to the news of his mother's death. It was Einstein's
favorite.

Three days starting Monday, March 15th were devoted to the
scientific part of the conference. The program on General
Relativity and Gravitational Physics was organized by Claus
Kiefer, the Chair of the Section, in consultation with other
office bearers. Plenary talks were given by Abhay Ashtekar
(Gravity, Geometry and the Quantum: Building on Einstein's
Legacy); Gerhard Huisken (Geometrical Concepts in General
Relativity); Hermann Nicolai (Cosmological Billiards); Asher Peres
(Quantum information and Relativity Theory) and Cliff Will (Was
Einstein Right?). In addition, there were several parallel
sessions on gravitational physics which featured invited talks by
Martin Bojowald (Quantum Cosmology); Christian Fleishhack (Loop
Quantum Gravity: Progress and Pitfalls); and Domenico Giulini (The
Thin-Sandwich problem: A Status Report), as well as a number of
contributed talks by others. In addition to scientific talks,
there was parallel session on ``Einstein and the Arts'' and a
plenary lecture by Arthur Miller, comparing Einstein's and
Picasso's views on space and time, in particular, simultaneity.

The city of Ulm had commissioned a special opera, {\it Einstein,
die Spuren des Lichts} (Einstein and the traces of light) from
Dirk D'Ase, the Composer in Residence in Vienna for this occasion,
with libretto by Joachim Stiller. The conference participants were
treated to a special performance on the evening of Tuesday, March
16th. The city has also organized a year-long exhibit to celebrate
the 125th birthday. There were several interesting items in the
exhibit. The one I found most moving was the desk that Einstein
used in the Berne Patent office ---it was just sitting there,
unprotected. One could touch it, even open the drawers. Alas,
there were no forgotten, left-over papers from 1905!

\parskip=10pt
\vfill
\eject

\section*{\centerline {
The 7th Eastern Gravity Meeting}}
\addcontentsline{toc}{subsubsection}{\it  
The 7th Eastern Gravity Meeting, by Deirdre Shoemaker}
\begin{center}
Deirdre Shoemaker, Cornell
\htmladdnormallink{deirdre-at-astro.cornell.edu}
{mailto:deirdre@astro.cornell.edu}
\end{center}

The seventh annual Eastern Gravity meeting took place on the lovely
campus of Bowdoin College in Brunswick, Maine from June 11-12, 2004.
Approximately thirty relativists participated in the conference.

In the first session, talks of the current status of evolving binary
black hole orbits in the fully-non-linear regime were
discussed. Gregory Cook (Wake Forest University) started the meeting
by relating the current state of efforts to generate
astrophysically realistic initial data for black-hole binaries.
In particular, he discussed the consequences of different boundary conditions
on the data.
Ulrich Sperhake (Penn State) presented evolutions using dynamical
excision with fixed mesh refinement and the BSSN-formulation.  He
pointed out that there are still difficulties in evolving a
boosted black hole with moving excision regions; nonetheless, he has
been able to evolve black holes with boosted velocities of 0.4c for
120M.  Wolfgang Tichy (Penn State) showed numerical simulations of
a black hole binary quasi-circular orbit also using BSSN and fixed mesh
refinement.  He emphasized the importance of using the gauge to keep
the apparent horizons in a fixed location long enough to evolve
one orbital time period.  Carlos Sopuerta (Penn
State) ended this session with a new effort to model
Extreme-Mass-Ratio Binaries numerically using finite elements.  
To date they have successfully completed a toy-model using scalar gravity.

The second morning session focused on binaries of a black-hole and a
companion star. The first talk was given by Monica Skoge (student at
Princeton) a former undergraduate student at Bowdoin.  She illustrated
a numerical method for the construction of quasi-equilibrium models of
black hole-neutron star binaries, concentrating the the construction
of such binaries in Newtonian gravity.  Thomas Baumgarte (Bowdoin
College) continued, focusing on the relativistic version of the
problem.  In their preliminary work, they located the onset of tidal
disruption in this fully relativistic framework in the
extreme-mass-ratio regime.  Pablo Laguna (Penn State) presented
gravitational waves from stellar disruptions by super-massive black
holes.  He showed that quadrupole gravitational waves emitted during
the tidal disruption process are described reasonable well (within
$\approx 5\%$) by a point particle approximation even in the strong
encounter case.  Finally Sasa Ratkovic (graduate student at SUNY at
Stony Brook) ended the session with a report on how different
equations of state for the companion (either quark or neutron star)
may result in a detectable difference in the gravitational wave
signatures.  These results are based on a pseudo-general relativistic
potential that incorporates post-Newtonian corrections.

The third session concerned mostly quantum cosmology.  Seth Major
(Hamilton College) started off with a discussion on the
consequences of a quantum cosmology arising from quantum geometry.
Building on the work of Martin Bojowald, he presented the solution
to the Lorentzian Hamiltonian constraint for
isotropic loop quantum cosmology coupled to a massless scalar
field. David Craig (Hamilton College) discussed
decoherent histories formulations of quantum theory 
and described applying these ideas to construct
a consistent quantum theory of recollapsing homogeneous universes,
the Bianchi IX cosmological models. Michael Pfenning (U.S. Military Academy) 
gave a brief introduction to the quantum weak energy inequalities
and showed ways the quantum inequalities can by used to constrain the 
magnitude of the Casimir vacuum energy density both above and below.
Daniel Cartin (Naval Academy
Preparatory School) followed recent work by Bojowald and others that
looked at cosmological models in terms of loop quantum gravity and
applied these methods to Bianchi I LRS spacetimes.

The final session of the day started with a discussion on self-force
effects on the ISCO of Schwarzschild by Steve Detweiler (University of
Florida). The self-force effects, from a scalar field, have been
calculated for a particle in a circular orbit of the Schwarzschild
geometry. Such effects change the radius and orbital frequency of the
innermost stable circular orbit.  Steve Drasco (graduate student at
Cornell) computed gravitational waveforms, and asymptotic fluxes of
energy and angular momentum produced by a spin-less point particle
moving along an arbitrary bound geodesic of a Kerr black hole accurate
to the first order in the mass ratio of the two bodies.  Etienne
Racine (graduate student at Cornell) presented a surface integral
derivation of post-1-Newtonian translational equations of motion for a
system of arbitrarily structured bodies, including the coupling to all
the bodies' mass and current multi-pole moments.

The second day of the conference started with Charles Evans
(University of North Carolina) presenting a means of specifying exact
outgoing-wave boundary conditions in time domain calculations with the
boundary at a finite distance from the isolated source. He applied
the method to 
the cases of a flat-space three-dimensional wave equation and 
Schwarzchild. Ian Morrison (undergraduate, Bowdoin
College) described the effects of differential rotation on the maximum
mass of Neutron Stars with nuclear equations of states.  He finds that
the maximum mass increases above the limit for non-rotating stars by
about 50$\%$.  David Garfinkle (Univ. of Guelph/Perimeter Institute)
showed the results for the numerical simulation of the approach to
the singularity in a general (no symmetry) vacuum spacetime, results
support the BKL conjecture; namely, as the singularity is approached spatial
derivatives become negligible and at each spatial point the dynamics
becomes that of a homogeneous, oscillatory spacetime. 
Steven Liebling (Long Island University)
discussed his use of non-gravitating, nonlinear models in three
dimensions using a distributed adaptive mesh refinement framework to
investigate the threshold behavior as a step in developing the
infrastructure to model the gravitational field equations.

The final session of the meeting began with Munawar Karim (St. John
Fisher College) who presented his work on a compact (10cm)
gravitational wave detector.  Eanna Flanagan
(Cornell) discussed how, when a source emits a burst of gravitational waves, 
different observers will measure different net changes in the angular 
momentum of the source, an effect related to the phenomenon of gravitational wave memory.
Douglas Sweetser presented a Rank 1 Unified Field Theory.  He contends that it
may be possible to quantize the 4D field equations using two spin
fields: spin 1 for EM and spin 2 for gravity.  R.L. Collins (retired,
The Univ. Texas at Austin) presented his case for a scalar alternate
to GR based on Mass-metric relativity.  He contends that the Gravity
Probe B will potentially invalidated one of these theories of gravity.

On behalf of the participants, congratulations again to Steve Drasco
for winning the Best Student Talk award.  All the students gave
outstanding talks, congratulations to each of them.  Our thanks to
Thomas Baumgarte and Bowdoin College for organizing a well run,
enjoyable meeting and reception.  Greg Cook has volunteered
to host the 8th Eastern Gravity meeting next year at Wake Forest
University in North Carolina, date to be announced.

\vfill
\eject

\section*{\centerline {
2004 Aspen winter conference on gravitational waves}
\\ \centerline{and their 
detection (GWADW)}}
\addcontentsline{toc}{subsubsection}{\it  
2004 Aspen GWADW, by Syd Meshkov}
\begin{center}
Syd Meshkov, Caltech
\htmladdnormallink{syd-at-ligo.caltech.edu}
{mailto:syd@ligo.caltech.edu}
\end{center}

The 2004 Aspen Winter Conference on Gravitational Waves and their
Detection (GWADW) was held at the Aspen Center for Physics, Aspen,
Colorado, Feb. 15-21, 2004.  The subtitle was ``Advancing Gravitational
Wave Detectors: Pushing the Quantum Limits". The special goal of this
workshop was to consider novel Gravitational Wave detectors beyond the
concepts used in Advanced LIGO and other future interferometers and
bars. Particular emphasis was placed on various methods that approach
and exceed the quantum limit, as, for example, with Quantum Non
Demolition (QND).

As usual for the Aspen conferences, each day of the workshop consisted
of about six hours of scheduled talks and discussion as well as
scheduled workshop interactions. The nature of the housing, with all
participants living and dining under the same roof, at the Aspen
Meadows, encouraged extensive opportunities for informal scientific
exchange. In addition, a public lecture was given on Wednesday,
February 18, 2004 at the Wheeler Opera House by Mark Coles (NSF),
entitled , ``The Universe... Live and in concert".

The opening session, chaired by Nergis Mavalvala and entitled,
``Pushing the Quantum Limits" , set the tone for the workshop. It
started with a talk by Nergis on ``Quantum Noise, Quantum Correlations
and the Standard Quantum Limit in GW Interferometers" . This was
followed by provocative talks by Yanbei Chen on ``Various Ways of
Beating the Standard Quantum Limit", and by T. Corbitt who discussed
``A Quantum noise simulation network". These talks stimulated lots of
discussion in the auditorium, on the slopes and at dinner.

Following a discussion by Richard Matzner on ``Constrained Evolution:
Concepts and New Results" , in a short session on Relevant
Astrophysical Sources, the subject of Techniques Addressing Quantum
Noise occupied the next few sessions. Ric DeSalvo discussed the
virtues of going underground in ``Mining for Gravitational Waves".
Incidentally, Going Underground is one of the two subtopics for the
2005 GWADW Meeting in Aspen. K. Somiya proposed an ``RF Readout scheme
to Overcome the SQL", In the first of two talks on Squeezed Light, R.
Schnabel discussed a ``Demonstration of Squeezed Light at Sideband
Frequencies below 100kH and T. Corbitt informed us about ``A
Ponderomotive Squeezed Vacuum Source".

Sessions on The Status of Existing Detectors followed. M. Cerdonio
told us about ``Wideband Operation of Upgraded Auriga', A. Gretarsson
reported on the staus of LIGO, B. Willke on the status of GEO 600, E.
Tournefier on the status of Virgo, and S. Sato reported both on the
status of TAMA and of a data taking run. A series of talks on next
generation detectors followed. Advanced LIGO was discussed by D.
Ottaway, and K. Kuroda informed us about the current staus of LCGT and
CLIO. B. Willke discussed Lasers for Advanced Interferometers and M.
Cerdonio reported on progress on the feasibility of DUAL. At this
point, late Wednesday, Feb. 18, Stan Whitcomb led a discussion on what
should drive the rest of the workshop. Such a discussion has become a
feature of this very interactive workshop, and resulted in the talks
that were eventually scheduled for the Summary session on Saturday
morning. Norna Robertson reported on the suspension design for
Advanced LIGO and R. DeSalvo discoursed on ``Bodies in Motion'.

A session on LISA followed. M. Tinto told us about a scheme for
optimal filtering of LISA data, R. Spero discussed ST7 Interferometer
Development, and M. te Plate described the LISA Test Package (LTP),
the Optical Bench system, and the LTP Inertial Sensor System.

Inasmuch as we are finally taking data, Stan Whitcomb reported on
results from LIGO and GEO Science runs. E. Majorana then talked on 
``Suspended Mirror Control: Learning through Virgo Experience". The next
few sessions were devoted to advances in critical areas, and to
lowering the sensitivity floor. These included some more LISA oriented
talks such as given by D. Shaddock, who talked about the LISA Optical
Bench, and a round table organized by A. Ruediger on ``One-Arm Locking
of LISA". The participants, in addition to Ruediger, were D. Shaddock,
I. Thorpe, and M. Tinto.  RSE was another important subject.
Beyersdork talked about implementing RSE via polarization control, S.
Kawamura talked about the current status of the 40m Detuned RSE
Prototype, and K. Somiya updated us on RSE in Japan.

In other talks, P. Beyersdorf talked about Technical Noise in QND and
R. Mittleman discussed progress in the important LIGO Livingston
Seismic Upgrade. M. Rakhmanov talked about how to measure the Dynamic
responses of the LIGO 4 km Fabry-Perot Arm Cavities.  Y. Aso discussed
the Suspension Point Interferometer.

The workshop concluded with a discussion of ``Important Questions"
raised during the Wednesday morning discussion led by Whitcomb. W.
Johnson explored the use of Bigger Masses and R. DeSalvo examined the
role of Gravity Gradients. F. Vetrano had the last word in his
discussion of the role of Atom Interferometry.

\vfill
\eject

\section*{\centerline {
Fifth LISA Symposium}}
\addcontentsline{toc}{subsubsection}{\it  
Fifth LISA Symposium, by Curt Cutler}
\begin{center}
Curt Cutler, Albert Einstein Institute
\htmladdnormallink{curt.cutler-at-aei.mpg.de}
{mailto:curt.cutler@aei.mpg.de}
\end{center}

LISA Symposia are held every two years, with the venue alternating
between the U.S. and Europe.  The latest one---the Fifth International
LISA Symposium---was held at ESTEC (The Netherlands) on July 12-15,
2004.  Oliver Jennrich was the main organizer, and it was attended by
about 180 scientists.

About 80 talks and almost 30 posters were presented 
over the course of this 4-day meeting.
Almost all the talks are now available on-line at

\htmladdnormallink
{http://www.rssd.esa.int/index.php?project=SP\&page=LISA\%20Symposium}
{http://www.rssd.esa.int/index.php?project=SP\&page=LISA\%20Symposium}.

The Proceedings will be published in a special issue of {\it Classical
and Quantum Gravity}.

The first day was devoted to LISA Pathfinder, a technology
demonstrator mission that will test key LISA technologies, especially
the inertial sensing and and drag-free control. Also, several
different kinds of thrusters (FEEPS, colloidal thusters, cold gas) 
will be employed, to help determine the best choice for
LISA.  LISA Pathfinder will carry two instruments: the European LISA
Test Package (LTP) and the U.S. Disturbance Reduction System (DRS).
Work on both sides of the Atlantic appears to be proceeding smoothly
towards LISA Pathfinder's scheduled launch in 2008.

Tuesday morning was devoted to status reviews from the
major ground-based detectors (both bars and interferometers).
The rest of the meeting was then devoted to sessions on 
LISA instrumental work, astrophysical sources, modeling and simulation,
and LISA data analysis.

The instrumental talks described significant
progress on many fronts, including the interferometry, test-mass 
charging and discharging, self-gravity gradients, and the thrusters.
The theoretical talks  mainly dealt with coalescences of 
massive ($\sim 10^6 M_{\odot})$ black holes,  inspirals of stellar-mass 
compact objects into
massive black holes (including the radiation reaction problem), 
and the problem of ``fitting out'' as many 
short-period galactic binaries as possible (important since confusion noise
from these binaries will actually dominate LISA's noise curve
in the low-to-mid-frequency range).



From the sessions on simulations and data analysis, a couple talks
were essentially advertisements for new LISA-related websites.  I'm
happy to use this space to help propagate that information.
M. Vallisneri and J. Armstrong have developed a software package
called {\it Synthetic LISA} for generating synthetic LISA time series,
i.e., for computing the TDI responses to gravitational waves and for
adding noise with the predicted spectrum. This is available at
\htmladdnormallink{http://www.vallis.org/syntheticlisa/}
{http://www.vallis.org/syntheticlisa/}.  A similar package called {\it The
LISA Simulator} has been built by N. Cornish, L. Rubbo, and
O. Poujade; it is available at 
\htmladdnormallink{http://www.physics.montana.edu/lisa/}
{http://www.physics.montana.edu/lisa/}.
Another relatively new site is the {\it Mock LISA Data Archive}
(
\htmladdnormallink
{http://astrogravs.nasa.gov/docs/mlda/}
{http://astrogravs.nasa.gov/docs/mlda/}), which contains a collection
of simulated data for different source classes, to be used in
developing and benchmarking data analysis algorithms.

I'll conclude by mentioning that the Symposium took place under a bit
of a cloud, politically. U.S. President Bush's new Vision for NASA
(announced in Jan.'04) involves a re-prioritization of NASA
activities, with more emphasis on manned missions (leading eventually to 
astronauts on Mars) and less on space science.  
To help free up money for the manned program, 
the budget for
the {\it Beyond Einstein} program was cut back significantly: the LISA
schedule was delayed by one year, Constellation-X was delayed two
years, and the Einstein Probes (like the Dark Energy Probe) were
eliminated from the budget completely. However, despite these
developments, knowledgeable insiders
generally opined that LISA enjoys sufficient support, both from Congress
and within NASA, that further budget tightening for science 
would probably just lead to further delays, as opposed to LISA's 
cancellation.

\vfill
\eject

\section*{\centerline {
GR17}}
\addcontentsline{toc}{subsubsection}{\it  
GR17, by Brien Nolan}
\begin{center}
Brien Nolan, Dublin City University
\htmladdnormallink{brien.nolan-at-dcu.ie}
{mailto:brien.nolan@dcu.ie}
\end{center}

GR17, the 17th International Conference on General Relativity and
Gravitation, took place in the Royal Dublin Society convention
centre, Dublin, Ireland from July 18th-23rd this summer. This
conference series, held under the auspices of the International
Society on General Relativity and Gravitation provides a forum for
a triennial review of all areas of research in gravitation, as
well as an opportunity to look forward to and plan for future
challenges. This report can only give a very brief flavour of the
content of the conference. The proceedings will be published next
year by World Scientific and in the meantime, abstracts and links
to some talks are available at the conference website
\htmladdnormallink{http://www.dcu.ie/\~{}nolanb/gr17.htm}
{http://www.dcu.ie/\~nolanb/gr17.htm} (please note that www.gr17.com will
not be in operation from late 2004 on). The conference was
attended by just under 700 scientific delegates.

The organization of the GRn conferences is in the hands of a
Scientific Organizing Committee, whose principal remit is to
select plenary speakers and workshop chairs, and a Local
Organizing Committee who handle organizational aspects. For GR17
these were chaired by Curt Cutler (AEI, Golm) and Petros Florides
(Trinity College Dublin) respectively.

The conference was opened on the morning of Monday 19th July by
Her Excellency Mrs Mary McAleese, President of the Republic of
Ireland, who delivered a well received address mentioning the
contributions of John Lighton Synge to relativity and William
Rowan Hamilton to physics generally. She also exhorted the
audience to make the most of the week's opportunities to learn and
discuss and to leave Dublin with a renewed commitment to their
vocation as scientists. Petros Florides presented Mrs McAleese
with copies of books by Stephen Hawking, Roger Penrose and Kip
Thorne to mark the occasion, which prompted her to comment that
she had never been given so much homework in her life. The opening
also included the presentation by the (outgoing) Society president
Bob Wald to Eanna Flanagan of the Xanthopoulous Award and a
memorial lecture on JL Synge by Petros Florides. Ted Newman spoke
briefly about the life and contribution to relativity of Peter
Bergmann, who passed away since the last GR meeting.

There were 16 plenary lectures and 19 workshops at GR17. The
former number represents a slight reduction on the number at GR16
held in Durban: this was done to allow more time for informal
discussion. The 19 workshops were divided between afternoon
parallel sessions and two evening poster sessions. The promise -
and subsequent appearance - of a glass or two of wine during the
poster sessions led to their being very well attended, with
discussions only dying down as RDS staff began turning off the
lights.

The SOC attracted an exceptional list of plenary speakers to GR17.
The scientific content of the conference was initiated by Sterl
Phinney (Caltech), who spoke about what we hope to learn about
astrophysics and relativity from the LISA mission, as well as what
subsequent missions may tell us about cosmology and the early
universe. Piotr Chru\'sciel (Tours) gave a review of key results
in Mathematical Relativity since GR16, focusing on global
properties of the (vacuum) Einstein equations. Don Marolf spoke
about the current status of the conjectured AdS/CFT correspondence
of string theory, as well as its more controversial
generalizations to other quantum theories of gravity.

Barry Barish (LIGO/Caltech) reviewed the status of the worldwide
network of gravitational wave observatories and the scientific
data currently being generated, and spoke about the prospects for
the future of gravitational wave detection by this network. John
Baez (UC Riverside) gave an expository talk on spin networks, spin
foams and loop quantum gravity, highlighting the theory's
successes in accounts of black hole entropy and the big bang.
Miguel Alcubierre (Mexico City) had the unenviable task of
reviewing recent progress in numerical relativity, both in terms
of the theoretical framework (hyperbolic formulations, gauge
choices, boundary conditions) and astrophysical simulations.

Wednesday morning saw the Gravity Research Funded session, chaired
by Louis Witten. GRF provides funding for each GRn conference to
invite speakers whose interests and expertise are in areas that
are tangential to or slightly outside general relativity. Thus the
audience at GR17 had the opportunity to hear Sir Martin Rees
(Cambridge) speak about black holes in active galactic nuclei, Jim
Peebles (Princeton) present a detailed critical review of the
$\Lambda-$CDM cosmological tests and John Preskill (Caltech)
introduce the theory of and prospects for quantum computing.

Preskill took to the stage again on Wednesday afternoon, when
Stephen Hawking (Cambridge) presented his proposed solution of the
black hole information paradox. Preskill jointly chaired the
session with Kip Thorne, who is the third party to a bet (Thorne
\& Hawking vs. Preskill) on the issue. The event, introduced by
Petros Florides, attracted a lot of media attention and so the
scientific audience was joined by numerous members of the fourth
estate to witness Hawking - but not Thorne - concede the bet and
to watch Preskill wave his prize (an encyclopedia of baseball,
from which information can be extracted at will) triumphantly over
his head. (This correspondent was greatly amused by media
reporting on the shenanigans, with references to an ``audience of
Dublin boffins" being stunned by Hawking's confession that he
``got the hole thing wrong". GR17 was referred to as the ``brains
Olympics" - in which case Preskill's parody of Olympian glory was
spot on.)

Back at the science end of things, Eric Poisson (Guelph) spoke
first on Thursday morning on the gravitational self-force (he
began by asking where all the TV cameras had gone). Licia Verde
(Pennsylvania) reviewed the implications for cosmology of the
first year WMAP data, and promised that the second year data would
be available "soon". Joseph Polchinski (UCSB) spoke about recent
work that shows that certain superstrings populating the early
universe could expand to cosmic scales today and be a significant
source of gravitational waves.

The first of the Friday morning lectures was given by Lars
Bildsten (KITP, UCSB), who spoke about recent advances in
relativistic astrophysics (sensitive X-ray astronomy; solar black
hole masses; data from PSR 0737-3039) and mentioned the consequent
challenges for theorists and the gravitational wave detection
community. Nergis Mavalvala (MIT) talked about the current status
of advanced gravitational wave detector technology, and how this
is now showing the way to a quantum-limited interferometer. The
final plenary lecture was given by David Langlois (IAP, Paris) who
reviewed the implications of the braneworld scenario for
gravitation and cosmology. The focus was on the standard
braneworld picture of our 4-d world resident in a 5-d bulk.

It is impossible to mention individually any of the large number
of oral and poster presentations made in the workshops, but note
that the conference book of abstracts is available online (see url
above). Close to 600 (different) abstracts were submitted in all,
amounting to some 400 talks and 200 posters. 
There were two public lectures held during the week. Kip Thorne
spoke on Monday night about what we will learn from the new
science of gravitational wave astronomy, and Roger Penrose spoke
on Friday night, giving his personal opinions on some popular
contemporary physical theories which he categorized under the
headings ``Fashion, Faith and Fantasy". Other business during the
week included the election of the new GRG Society officers; Cliff
Will succeeds Bob Wald as president.

Astonishingly for Dublin, delegates encountered only one day of
rain, allowing us to have lunch and coffee in the RDS grounds most
days and enjoy what passes for good weather here. We look forward
to even more sunshine at GR18 in Sydney in 2007. 

\vfill
\eject

\section*{\centerline {
Loops and Spinfoams}}
\addcontentsline{toc}{subsubsection}{\it  
Loops and Spinfoams, by Carlo Rovelli}
\begin{center}
Carlo Rovelli, CNRS Marseille
\htmladdnormallink{rovelli-at-cpt.univ-mrs.fr}
{mailto:rovelli@cpt.univ-mrs.fr}
\end{center}

The ``Loops and Spinfoams" conference was held in
Luminy, Marseille, France, on May 3 to 7, 2004.  Aim of the
conference was to make the point on the loop approach to
quantum gravity.

Participation has been much higher than expected: more than
a hundred participant, including a large number of young
researchers and students.  Discussion has been intense and
the atmosphere friendly and very lively.

The conference was articulated in thematic days, focused on:
(i) Canonical loop quantum gravity, (ii) the spinfoam
formalism, (iii) applications, and (iv) related approaches. 
The program was developed in 45 presentations of different
lengths, too many for illustrating them here individually. 
Very ample time was left for questions and for discussion, a
format that has proven effective and has been appreciated by
the participants.

The overall picture of the research in nonperturbative
quantum gravity that has emerged is encouraging.  The loop
approach, with its rich variety of versions and formalisms,
is still incomplete, and a large number of issues remain
open.  But there is a large common ground in the variety of
points of view.  A general understanding on how to formulate
background independent quantum field theory exists and
yields to a credible hypothesis of solution to the quantum
gravity puzzle.  In addition, loop quantum gravity is
finding an increasing spectrum of applications.

A general discussion during the last morning has been based
on a list of questions proposed by the audience.  The list
of these questions (quite interesting, and requested by
many) can be found from my home page 
\htmladdnormallink
{http://www.cpt.univ-mrs.fr/\~{}rovelli}
{http://www.cpt.univ-mrs.fr/\~rovelli}.  
Pictures of the conference
can be found in
\htmladdnormallink
{http://perimeterinstitute.ca/activities/scientific/cws/marseille.cfm}
{http://perimeterinstitute.ca/activities/scientific/cws/marseille.cfm}.
Publication of proceedings has been suggested and is being
considered.

The weather, uncharacteristically grey during the first
days, has cleared up for the free afternoon, allowing the
participants to spread along the sea and in the marvelous
wilderness of Marseille's ``Calanques".

\vfill
\eject

\section*{\centerline {
20th Pacific coast gravity meeting}}
\addcontentsline{toc}{subsubsection}{\it  
20th Pacific coast gravity meeting, by Michele Vallisneri}
\begin{center}
Michele Vallisneri, JPL
\htmladdnormallink{vallis-at-caltech.edu}
{mailto:vallis@caltech.edu}
\end{center}

The 20th Pacific Coast Gravity Meeting (PCGM20) was held at the
California Institute of Technology, Pasadena, on March 26 and 27,
2004. Faithful to its tradition, this was an informal and relaxed
regional meeting where all participants who wished to speak had a
chance to engage a welcoming audience for all of twelve intense
minutes. As always, the aim was to foster communication and
understanding among gravitational physicists with different
backgrounds.  The conference was organized by the writer of this
report with the invaluable help of JoAnn Boyd and Yanbei Chen, and
with the wise and benign oversight of Kip Thorne.

This year's meeting was graced by the enthusiastic participation of
more than 80 researchers, 45 of which gave talks. The first day began
with a session on gravitational-wave astrophysics, with Shane L.
Larson, Rick Jenet, Rafael Araya-Gochez, Marc Favata, Sherry Suyu, and
Geoffrey Lovelace; the next session, on ground-based interferometers,
was animated by Patrick Sutton, Jan Harms, Yi Pan, Akira Villar, Peter
Shawhan, and Szabolcs Marka. Vladimir Braginsky opened the afternoon
with a charming talk on the ``Adolescent Years of Experimental
Physics;'' he was followed by more theoretical talks, by Alfonso
Agnew, Ivan Avramidi, William Pezzaglia, Jack Hohner, Steve Giddings,
Belkis Cabrera-Palmer, Henriette Elvang, Keith Copsey, and James
Dunham. In the evening, the weary participants found refreshment(s) at
an animated party at Kip Thorne's house.

The second day began on (or near) cosmology, with talks by Zoltan
Perjes, Dominic Clancy, Jim Isenberg, Lior Burko, and Donald Marolf;
it continued on research relevant to LISA, with Teviet Creighton,
Naoki Seto, Daniel Bambeck, Seth Timpano, Louis Rubbo, and Jeff
Crowder. The afternoon sessions dwelt on numerical relativity, with
Harald Pfeiffer, Mark Scheel, Robert Owen, Ilya Mandel, Luisa Buchman,
and Frans Pretorius; and the meeting closed with Mihai Bondarescu
(delivering Pavlin Savov's talk), Richard Price, David Meier, Craig
Hogan, Gary Horowitz, and Martin Kaplan. The titles of all talks can
be seen on the PCGM20 website, www.tapir.caltech.edu/pcgm20.

Fifteen students competed for the award for the best student
presentation, previously known as the Jocelyn Bell Burnell prize, and
now sponsored by the APS Topical Group on Gravitation, which was
awarded jointly to Henriette Elvang (University of California, Santa
Barbara), and Louis Rubbo (Montana State University, Bozeman).

The next PCGM will be held at the University of Oregon, Eugene, most 
probably on March 25 and 26, 2005. Watch out for the announcement.

\end{document}